# 4MOST Consortium Survey 9: One Thousand and One Magellanic Fields (1001MC)

Maria-Rosa L. Cioni[1]
Jesper Storm[1]
Cameron P. M. Bell[1]
Bertrand Lemasle[2]
Florian Niederhofer[1]
Joachim M. Bestenlehner[3]
Dalal El Youssoufi[1]
Sofia Feltzing[4]
Carlos González-Fernández[5]
Eva K. Grebel[2]
David Hobbs[4]
Mike Irwin[5]
Pascale Jablonka[6]
Andreas Koch[2]
Olivier Schnurr[1,7]
Thomas Schmidt[1]
Matthias Steinmetz[1]

[1] Leibniz-Institut für Astrophysik Potsdam (AIP), Germany
[2] Zentrum für Astronomie der Universität Heidelberg/Astronomisches Rechen-Institut, Germany
[3] Physics and Astronomy, University of Sheffield, UK
[4] Lund Observatory, Lund University, Sweden
[5] Institute of Astronomy, University of Cambridge, UK
[6] Laboratoire d'astrophysique, École Polytechnique Fédérale de Lausanne, Switzerland
[7] Cherenkov Telescope Array Observatory, Bologna, Italy

The One Thousand and One Magellanic Fields (1001MC) survey aims to measure the kinematics and elemental abundances of many different stellar populations that sample the history of formation and interaction of the Magellanic Clouds. The survey will collect spectra of about half a million stars with $G < 19.5$ magnitudes (Vega) distributed over an area of about 1000 square degrees and will provide an invaluable dataset for a wide range of scientific applications.

## Scientific context

During the last decade, our view of the Magellanic Clouds has changed significantly. These galaxies most likely approached the Milky Way only a few Gyr ago, rather than having orbited around it for a Hubble time (for example, Kallivayalil et al., 2013). This has motivated many studies aimed at explaining the structure and the star formation history of the Large Magellanic Cloud (LMC), the Small Magellanic Cloud (SMC) and their tidal features, i.e., the Magellanic Bridge and Stream. Furthermore, with the increased number of deep imaging observations we have discovered potential satellite galaxies of the Magellanic Clouds and new stellar streams possibly associated with tidal stripping events (for example, Koposov et al., 2018).

The Magellanic Clouds are the largest and most massive satellite galaxies of the Milky Way. The LMC resembles a spiral galaxy, with a rotating disc, an off-centre bar, and a few spiral arms. Young, intermediate-age and old stars show different levels of substructures extending to large radii. The LMC hosts the most massive stars known today (for example, Bestenlehner et al., 2014). The SMC is a dwarf spheroidal galaxy, with a significant depth along the line of sight and a morphology shaped by tidal interactions (for example, Niederhofer et al., 2018). The SMC formed half of its stellar mass prior to an age of ~ 6 Gyr (for example, Rubele et al., 2018). The Magellanic Bridge is the product of an LMC–SMC collision ~ 200 Myr ago; it is likely formed of SMC material and it contains both gas and stars. The origin of the Magellanic Stream, which is made of both LMC and SMC gas (for example, Richter et al., 2013), depends on the orbital history of the Magellanic Clouds and their one-to-many interactions with the Milky Way and with each other.

Large amounts of telescope time have been invested in imaging the Magellanic Cloud stars, studying their distribution, and measuring their ages, distances, and motions. A major ESO programme, that will provide targets for spectroscopic follow-up studies, is the VISTA survey of the Magellanic Clouds system (VMC[1]), aimed at deriving the spatially resolved star formation history and three-dimensional geometry of the system. The VMC is the most sensitive high-spatial-resolution survey of the Magellanic Clouds in the near-infrared to date. The VMC and other contemporary surveys in the optical domain, such as the SMC in Time: Evolution of a Prototype interacting late-type dwarf galaxy (STEP), the Survey of the MAgellanic Stellar History (SMASH), and the Optical Gravitational Lensing Experiment (OGLE), were made possible by the development of wide-field cameras at telescopes dedicated to survey observations for a large fraction of the available time.

Next to this wealth of photometric observations, which have yet to reach their full exploitation (also including data from the Gaia satellite), there is a pronounced lack of spectroscopic observations across the range of stellar populations and substructures of the Magellanic Clouds. The largest samples of moderate resolution (at least $R = 4000$) spectra, suitable for kinematics and metallicity measurements, comprise about 9000 giant stars (for example, Dobbie et al., 2014). Chemical tagging, a powerful tool to discern the history of stellar populations, requires high-resolution (at least $R = 20\,000$) spectra and these only exist for about 4000 giant stars and 1300 early-type massive stars, and for much smaller samples of other stellar types (for example, Nidever et al. 2019). Despite the major scientific advances of these programmes, where two thirds of the spectra refer to LMC stars and one third to SMC stars, they are far from providing a comprehensive view of a system where stars span the age range of the Universe and that is strongly shaped by dynamical interactions.

## Specific scientific goals

The 1001MC survey aims are as follows:
– To find and characterise kinematic and chemical patterns within the Magellanic Clouds system.
– To study links between kinematics and chemical patterns as well as their spatial distribution across different stellar populations.
– To establish how the star formation history and the dynamical evolution of the system are related to these patterns.
– To study the metallicity-dependent physical and wind properties of massive stars and their evolutionary stages.
– To quantify the metallicity dependence of key distance indicators.

A comprehensive study of the kinematics and chemistry of a large number of stars at different evolutionary phases and with a wide spatial distribution is needed to address these goals.

Line-of-sight (radial) velocities are one of the fundamental components of motion to describe the internal kinematics of galaxies from which the distribution of mass is estimated. Radial velocities, together with proper motions, are necessary to derive space velocities from which to infer orbital motions. The 1001MC survey will provide radial velocities that match the accuracy of the tangential velocities (proper motions) that are measured using astrometry from, for example, the VMC survey (Niederhofer et al., 2018) and Gaia. These are of the order of 2.5 km s$^{-1}$ for an ensemble of stars (which corresponds to 1% of the radial velocity of the Magellanic Clouds and is a factor of 10 smaller than the internal motion), considerably improving our capability to spatially sample kinematical substructures within the Magellanic Clouds.

The iron abundance [Fe/H] is usually used as a proxy for the metallicity of stars. Age and metallicity of red giant branch stars are, however, degenerate in the colour-magnitude diagrams. Moreover, the dependence on metallicity of key distance indicators (for example, the period-luminosity relations of Cepheids, the luminosity of red clump stars) is still not well assessed. The 1001MC survey will derive the metallicity of different stellar populations, provide a metallicity map of the system as a function of age, and lift some of the degeneracies that affect the tracers of stellar population parameters. For bright stars, in addition to their radial velocity and iron abundance, we will measure the abundances of several α-elements, Fe-peak elements, and elements produced in the slow (s-) and rapid (r-) neutron-capture nucleosynthesis processes (for example, Zr, Ba, Sr, Eu), elements that will provide further constraints on the chemical enrichment history of the different components of the two galaxies. We will also classify stars throughout the Hertzsprung-Russell diagram from their spectral features. In particular we will obtain a complete census of massive stars (> 15 $M_\odot$) including O main-sequence stars, blue, yellow, and red supergiants and Wolf-Rayet stars. We aim also to provide some indication of the binary nature of these stars from radial velocity variations (for example, Sana et al., 2013) and derive approximate systemic velocities of variable stars using templates, for example in the case of Cepheids, RR Lyrae stars, and long-period variables (for example, Nicholls et al., 2010).

The 1001MC survey data will also be used to quantify and map the dust absorption within the Magellanic Clouds using background galaxies. This is achieved by comparing the rest-frame spectra of galaxies with a reddening free template where the adjustment of the continuum level will provide a measure of the dust content along the line of sight (for example, Dutra et al., 2001). Compared to an ongoing study, based on the analysis of spectral energy distributions, spectra provide the redshifts of galaxies which are needed to scale the templates, reducing considerably the uncertainties associated with a photometric determination. We estimate that with 120 galaxies per square degree we would obtain a dust map with a spatial resolution of 0.143 square degrees, directly comparable to that of current star formation history studies (for example, Rubele et al., 2018).

### Science requirements

The 1001MC survey aims to reach accuracies of ± 2 km s$^{-1}$ for the radial velocities of individual stars. This accuracy is designed to match the accuracy of the proper motion obtained with other facilities. For example, for individual bright stars in the Magellanic Clouds Gaia will provide proper motion accuracies of 2.5 km s$^{-1}$ or obtain similar accuracies on bulk velocities from the average of about 200 G-type or 3000 M-type stars. These values have been calculated assuming the Gaia end-of-mission proper motion accuracies[2] and considering that at the distance of the Magellanic Clouds 0.01 milliarcseconds per year is roughly equivalent to 2.5 km s$^{-1}$.

We also aim to obtain metallicities for individual stars with accuracies better than ± 0.2 dex. This will be achieved using Fe lines and/or indirectly using the Ca II and Mg b triplets. Stars in the Magellanic Clouds span a relatively large range of metallicities, from about half solar to very metal-poor (Δ[Fe/H] > 2 dex), and uncertainties in the Fe abundance of 0.2 dex or better are needed to distinguish between different stellar populations. These can be derived from spectra of individual stars, but for some faint targets the spectra will be combined to reach a minimum signal-to-noise ratio (S/N) of 20 per Å within the spectral regions above. We aim to reach similar accuracies in measuring the abundance of other prominent elements.

A fraction of the 1001MC targets are variable stars, which means that they change temperature across the pulsation cycle. To derive elemental abundances for Cepheids it is necessary to acquire spectra before the temperature changes significantly and therefore we plan to observe for 1 hour at any given epoch, this time being sufficient to reach the required S/N. RR Lyrae stars, instead, are faint and for them we will focus on kinematics. However, we will explore the possibility of also measuring average Fe abundances by combining, for example, stars with similar metallicities as estimated from the Fourier decomposition of their light-curves.

### Target selection and survey area

The 1001MC survey will cover an area of about 1000 square degrees (Figure 1). This area comprises targets that trace the extent of different stellar populations and that describe substructures throughout the Magellanic Clouds. The 1001MC survey will obtain 4MOST spectra of about half a million stars with G < 19.5 magnitudes and produce a sample that is a factor of 20 larger than the largest sample of Magellanic Cloud stellar spectra assembled in the past. Spectra from the Gaia Radial Velocity Spectrograph reach bright red and blue supergiant stars, while completed and ongoing observing campaigns with 2dF and APOGEE-2S observe giant stars well above the red clump in the Magellanic Clouds. The 1001MC observations will prioritise areas with the highest legacy value (i.e., the central LMC and SMC areas).

The 1001MC targets span a range of about 10 magnitudes (Table 1, Figure 2) and comprise young massive and supergiant stars, intermediate-age giant stars (asymptotic giant branch stars of M and C type, red clump stars), old red giant and horizontal branch stars. They also include different types of variable stars (Cepheids of any type, long-period variables, and RR Lyrae stars). The number of targets reflects the need to statistically characterise and spatially trace substructures of the Magellanic Clouds by age, chemical content and, if possible, multiple diagnostics.

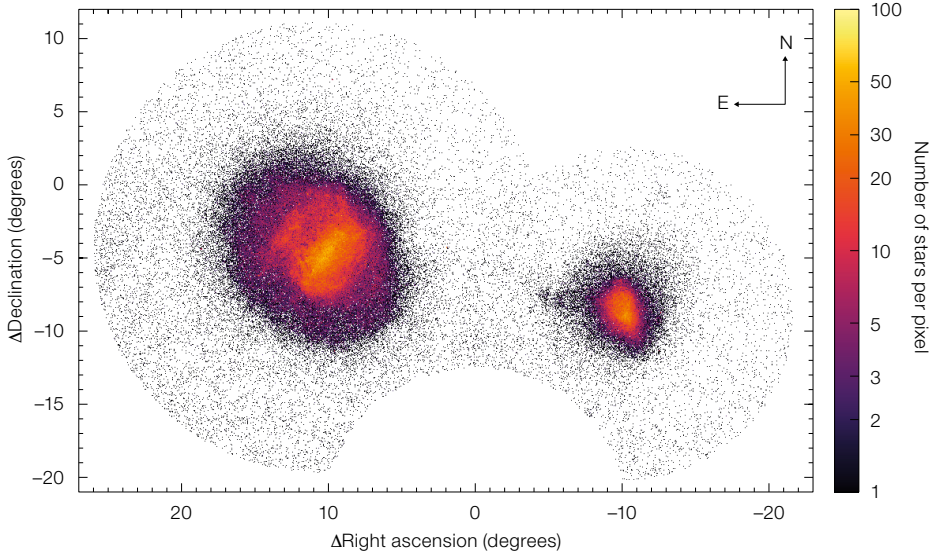

Figure 1. Spatial source density distribution of the 1001MC stars in a zenithal equidistant projection with the origin at a right ascension of 50 degrees and declination of −67.5 degrees. Pixels are 0.05 square degrees. The total area covered results from the combination of two circular regions, each centred on one of the Magellanic Clouds, that encompass their extended structure and possible tidal features. The circular cutout in the coverage at the bottom of the figure is caused by a declination > −80 degree selection limit, which was implemented because observing at lower declinations becomes really inefficient with 4MOST.

The selection of 1001MC targets results from the combination of near-infrared observations from VISTA and 2MASS with optical observations from Gaia, where variable stars are identified in OGLE and Gaia data. In particular, Gaia parallaxes are used to remove Milky Way stars and to create a catalogue with homogeneous coordinates. The VISTA selection data originate from the VMC survey for the central regions of the Magellanic Clouds and from the VISTA Hemisphere Survey (VHS) for the outer regions. VISTA data will be used to select targets for the Low-Resolution Spectrograph (LRS) while targets for the High-Resolution Spectrograph (HRS) will be selected from 2MASS. Several sub-surveys are defined to represent the range of stellar populations within the Magellanic Clouds. Table 1 shows the approximate number of targets that belong to each sub-survey while Figure 2 shows the magnitude and colour criteria adopted to select the targets.

The central regions of the Magellanic Clouds (a few hundred square degrees) will be observed more frequently (3–6 visits) because of the high target density. This will then allow us to monitor a subset of stars across the different types. Some stars will be monitored with the HRS and others with the LRS. The main goal of these monitoring campaigns is to trace the variation in the radial velocity curves that are directly linked to the internal structure of pulsating stars and/or to the presence of companions, as well as to establish the effect of binaries on the dynamics of the different stellar systems. In addition, we will make use of the poor observing conditions programme (Guiglion et al., p. 21) for those targets that fall within the 1001MC area.

### Spectral success criteria

To meet the goals of the 1001MC survey we require spectra with a S/N per Å of about 100–1500 encompassing both LRS and HRS observations of targets distributed across the Hertzsprung-Russell diagram. Depending on stellar magnitude and type of star, we will measure elemental abundances and/or radial velocities. In practice, for the abundance of specific elements we will use spectra with S/N > 80–100 per Å, while for determining indirect Fe abundances with either the Ca II and the Mg b triplets we will use spectra with S/N > 20–30 per Å. From all of the other spectra of individual stars with S/N < 20 per Å we will derive radial velocities. Furthermore, we will combine spectra of individual stars, for example RR Lyrae stars, to estimate median element abundances and radial velocity.

The S/N is measured in the continuum within one of the following spectral ranges: 5140–5200 Å (including the Mg b triplet at: 5167, 5172, 5183 Å) and 8350–8850 Å (including the Ca II triplet at: 8498, 8542, and 8662 Å). In this way there will be at least one spectral region with sufficient S/N to derive the radial velocity and some elemental abundance(s) of each star.

Table 1. Sub-surveys and preliminary numbers of 1001MC targets. For the main sequence, red clump giant and red giant branch stars, these numbers represent only fractions of the respective populations (Figure 2; LRS only): 15% for main sequence stars, 5% for red clump stars, and 10% for red giant branch stars.

| Star type | Spectrograph | Magnitude range | Number of targets (10³) |
| --- | --- | --- | --- |
| Main sequence stars | LRS | 11.9 < G < 19.5 | 87 |
| Red clump giant stars | LRS | 16.9 < G < 19.5 | 64 |
| Red giant branch (RGB) stars | LRS, HRS | 14.1 < G < 19.5 | 310 |
| RR Lyrae stars | LRS | 12.2 < G < 19.5 | 36 |
| Cepheids | HRS, LRS | 11.5 < G < 19.5 | 9 |
| Supergiant stars | HRS, LRS | 10.5 < G < 19.5 | 36 |
| O-rich asymptotic giant branch (AGB) stars | HRS | 10.8 < G < 19.5 | 25 |
| Carbon stars | HRS | 12.4 < G < 19.5 | 9 |
| Background galaxies | LRS | | 120 |

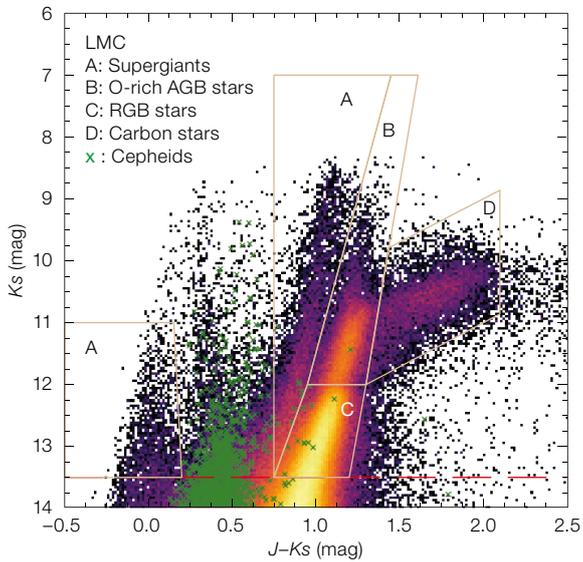 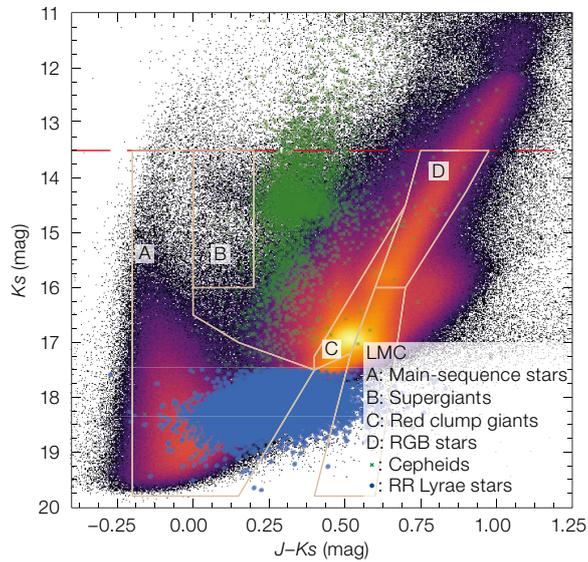

Figure 2. Near-infrared, $J–K_s$ versus $K_s$, colour-magnitude Hess diagrams of the 1001MC stars across the LMC (targets in the SMC are ~ 0.5 magnitudes fainter). The red dashed lines indicate the separations between the HRS targets (left) and the LRS targets (right), except for Cepheids. All sources have $G < 19.5$ magnitudes except for the RR Lyrae stars that extend to fainter magnitudes.

To estimate the progress of the 1001MC survey we define a figure of merit (FoM). The FoM for each sub-survey is the ratio of observed targets to the goal number, and the FoM for the whole survey corresponds to the minimum FoM among the sub-surveys. In this way each stellar population will be sufficiently represented across the Magellanic Clouds system.

Acknowledgements

We acknowledge funding from the European Research Council (ERC) under the European Union's Horizon 2020 research and innovation programme (grant agreement No 682115 as well as from the German Research Foundation (DFG) via Sonderforschungsbereich "The Milky Way System" (SFB 881).

Links

[1] VISTA survey of the Magellanic Clouds system (VMC): http://star.herts.ac.uk/~mcioni/vmc/
[2] Expected nominal science performance of the Gaia mission: https://www.cosmos.esa.int/web/gaia/science-performance

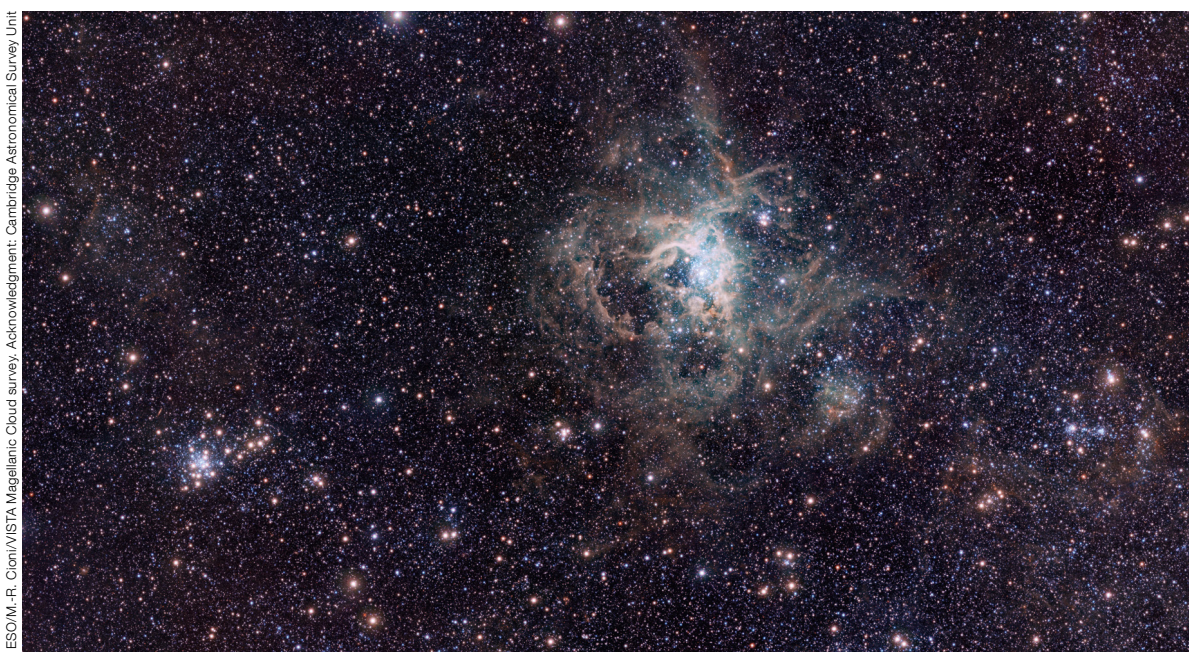

This VISTA image of the 30 Doradus star-forming region (or Tarantula Nebula) from the VMC Public Survey is being used to study the detailed star formation history of the Magellanic system.

ESO/M.-R. Cioni/VISTA Magellanic Cloud survey. Acknowledgment: Cambridge Astronomical Survey Unit